# Production and decay of *K*-shell hollow krypton in collisions with 52 – 197 MeV/u bare xenon ions

Caojie Shao,[1, 2, 3] Deyang Yu,[1, †] Xiaohong Cai,[1, ‡] Xi Chen,[4] Kun Ma,[5] Jarah Evslin[1], Yingli Xue,[1] Wei Wang,[1] Yury. S. Kozhedub,[6] Rongchun Lu,[1] Zhangyong Song,[1] Mingwu Zhang,[1] Junliang Liu,[1, 3] Bian Yang,[1] Yipan Guo,[1, 3] Jianming Zhang,[1, 3] Fangfang Ruan,[1] Yehong Wu,[1, 3] Yuezhao Zhang,[1, 3] Chenzhong Dong,[4] Ximeng Chen,[2] and Zhihu Yang[1]

[1]*Institute of Modern Physics, Chinese Academy of Sciences, Lanzhou 730000, China*
[2]*School of Nuclear Science and Technology, Lanzhou University, Lanzhou 730000, China*
[3]*University of Chinese Academy of Sciences, Beijing 100049, China*
[4]*College of Physics and Electronic Engineering, Northwest Normal University, Lanzhou 730070, China*
[5]*School of information engineering, Huangshan University, Huangshan 245041, China*
[6]*Department of Physics, St. Petersburg State University, St. Petersburg 198504, Russia*

([†]Email: d.yu@impcas.ac.cn; [‡]Email: caixh@impcas.ac.cn; *Accepted by Physical Review A*)

X-ray spectra of *K*-shell hollow krypton atoms produced in single collisions with 52 – 197 MeV/u $Xe^{54+}$ ions are measured in a heavy-ion storage ring equipped with an internal gas-jet target. Energy shifts of the $K\alpha_{1,2}^{s}$, $K\alpha_{1,2}^{h,s}$, and $K\beta_{1,3}^{s}$ transitions are obtained. Thus, the average number of the spectator *L*-vacancies presented during the x-ray emission is deduced. From the relative intensities of the $K\alpha_{1,2}^{s}$ and $K\alpha_{1,2}^{h,s}$ transitions, the ratio of *K*-shell hollow krypton to singly *K*-shell ionized atoms is determined to be 14 – 24%. In the considered collisions, the *K*-vacancies are mainly created by the direct ionization which cannot be calculated within the perturbation descriptions. The experimental results are compared with a relativistic coupled channel calculation performed within the independent particle approximation.

## I. INTRODUCTION

*K*-shell hollow atoms, i.e., atoms with both *K*-shell electrons removed while the outer shells remain occupied, are very special atomic species, which exist under conditions which are far from equilibrium [1,2]. Heavy hollow atoms provide a unique opportunity to study angular momentum coupling and electron-electron correlation in an exotic regime, where the Breit interaction and relativistic effects play a more pronounced role than in light atoms [3-5]. They are also ideal medium for investigating exotic decay modes, such as the hypersatellite transitions [6], the hypersatellite Auger process [7,8], the two-electron one-photon (TEOP) transitions [9], the three-electron Auger process [10], as well as the dynamics of violent collisions [11,12]. In addition to their interest for fundamental atomic physics, production mechanisms and decay properties of such atoms are also important for high energy density plasma [13,14], hard x-ray laser [15], and molecule imaging [16,17] research.

*K*-shell hollow manganese atoms, resulting from the *K*-electron capture decay of $^{55}$Fe, were first observed by Charpak *et al.* [18]. Later, the x-ray coming from the $K^{-2} \rightarrow K^{-1}L^{-1}$ Gallium transition was observed in the *K*-electron capture decay of $^{71}$Ge by Briand *et al.* [6], and named as the *hypersatellite line* because its energy is much more shifted than the $K^{-1}L^{-1} \rightarrow K^{0}L^{-2}$ satellite transitions from the normal diagram lines. Thereafter, energetic electrons [19,20], light ions [21-23], and photons [20,24-31] have been employed in collisions with solid targets, in order to create double *K*-shell vacancies. *K*-shell hollow atoms produced in these collisions are beyond radioactive isotopes. In both methods, the first *K*-shell electron is either captured by the radioactive nuclei or ejected by the projectile, while the second is generally shaken off due to the sudden change of nuclear charge or the electron-electron correlation. Because of the small shake-off probability, the yield of *K*-shell hollow atoms relative to the single *K*-vacancy atoms is usually as low as $10^{-6} - 10^{-2}$, depending on the nuclear charge [32]. The low intrinsic cross sections require a very high beam flux as well as a dense target in order to obtain reasonable count rates for the emitted hypersatellite x-ray or Auger electrons from *K*-shell hollow atoms.

In collisions between energetic heavy ions and atoms, the two *K*-shell electrons of the target atom can be removed dominantly by two independent direct single-ionization events [33] rather than by a correlation-mediated shake-off following a single-ionization event, and therefore double-*K*-vacancies can be created with a much higher probability. Using a 30 MeV $O^{5+}$ beam, Richard *et al.* [34] directly measured the $K\alpha$ hypersatellites from calcium, finding a relative intensity close to 0.5% with respect to the satellite lines. Thereafter, many kinds of heavy ions, but usually lighter than the target and with several MeV/u energies, have been employed to bombard various targets in order to produce double *K*-shell vacancies. In these cases, it has been found that the cross-section ratio of double-to-single *K*-shell vacancies, $R_{21} = \sigma_{K^{-2}}/\sigma_{K^{-1}}$, is only a few percent, and roughly proportional to $Z_{\mathrm{P}}^{2}$, where $Z_{\mathrm{P}}$ is the nuclear charge of the projectile [35,36]. Furthermore, a higher ratio $R_{21}$ can be achieved when heavier ions are employed. The $R_{21}$ generally reaches its maximum value when the collision velocity is close to the classical velocity of the target *K*-shell electron. In particular, when bare heavy ions are utilized, a much higher ratio $R_{21}$ of about 20 – 36% has been observed due to the contribution of the *K*-shell-to-*K*-shell electron transfer [37-39]. Moreover, hollow atoms can also be produced with a large yield during the electron pickup by slow highly charged ions (HCI) from a solid surface [7,10,40,41]. When bare ions are employed, *K*-shell hollow atoms can be created in this way [42]. It should be noted that, although in this method the hollow atoms are produced above or at the surface, the decay mainly occurs below the surface due to the image-charge acceleration effect [1,2].

The yields of the *K*-shell hollow atoms, produced from a solid target bombarded by either an energetic heavy or a slow bare ion beam, are sufficient to perform a precise x-ray or Auger spectroscopy measurement. However, a number of investigations show that both creation and decay processes of hollow atoms are significantly affected by environmental effects [43-46]. First, a target atom may collide with a secondary particle outgoing from another target atom, rather than with a primary projectile. Second, interactions between neighboring atoms, e.g., interatomic transitions, will interfere with the rearrangement of the primary vacancies before a measurable decay occurs. Production of a large quantity of *free* heavy *K*-shell hollow atoms from single collisions of isolated target atoms would be highly advantageous for precise x-ray or Auger spectroscopy studies. This will allow us to compare the results with theories in which isolated atoms are usually employed to model the physical processes. A significant step forward has been made by employing a 2.1 keV/u $N^{6+}$ beam passing through a thin nickel microcapillary foil, where free hollow atoms were first extracted in vacuum [47]. However, this technique is limited to such long-lifetime hollow atoms, rather than *K*-shell hollow, heavy ones [47,48].

In recent years, the tremendous progress in free electron lasers (FELs) has opened up a new way to produce free *K*-shell hollow atoms. The high-intensity x-ray pulses allow the two *K*-shell electrons of an atom to be removed dominantly through a sequential single-photon photoionization process within the core-vacancy lifetime. The production cross section of *K*-shell hollow atoms for this process is greatly increased as compared with single-photon double ionization process. The creation of neon *K*-shell hollow atoms at the Linac Coherent Light Source (LCLS) was first demonstrated by detecting hypersatellite Auger electron spectra [49]. The yield of the hypersatellite lines was 30 times larger compared to earlier synchrotron experiments. Later, various light *K*-shell hollow atoms or molecules (e.g., $N_2$ and $CO_2$) were also created by the FELs [50-52]. In case of medium- to high-*Z* atoms, the high binding energy of the *K*-shell electrons and short lifetime of single *K*-shell ionized atom require a higher photon energy and intensity in order to improve the cross-section ratio of double-to-single *K*-shell vacancies. The $K\alpha^h$-to-$K\alpha^s$ ratio is about $3.95\times10^{-4}$ for krypton for the case of an x-ray pulse with an equivalent density of $6.3-7.1\times10^{32}$ photons/cm$^2$s and energy of 15 keV [53].

In addition to FELs, energetic heavy ion beams colliding with gaseous targets are alternative tools that can be employed to produce free *K*-shell hollow atoms with high production yields, and can provide useful information on the collision mechanisms as well as on the atomic structure. Thereby, the production of *K*-shell hollow atoms strongly depends on the perturbation strength from projectile ion, $\kappa = Z_\mathrm{P}/v_\mathrm{P}$, where $Z_\mathrm{P}$ is the nuclear charge of the ion and $v_\mathrm{P}$ is its velocity in atomic units. A few experiments have been carried out to produce free *K*-shell-hollow atom by using energetic bare ions. For examples, hollow lithium atoms have been created in collisions with $N^{7+}$ at 10.6 MeV/u ($\kappa = 0.34$) [54] and $Ar^{18+}$ at 95 MeV/u ($\kappa = 0.31$) [55,56], with the ratio $R_{21}$ of only 0.36% and 2.3%, respectively. In these cases, the *K*-vacancy producing mechanisms are mainly ionization and excitation channels. A higher relative yield of 36% *K*-hollow argon atoms was observed in collisions with $Fe^{26+}$ at 7.7 MeV/u ($\kappa = 1.54$) [39], as a result of strong *K*-shell-to-*K*-shell electron transfer since the velocity of the ion is comparable with the *K* orbital velocity of the target atom.

However, no experiment has been reported so far for double *K*-shell ionization of heavier target atoms ($Z_\mathrm{T} \geq 30$, $Z_\mathrm{T}$ is the nuclear charge of the target atom) colliding with much heavier bare ions at high energies. In this case, a high relative yield of *K*-hollow target atoms is expected according to the very high probability of direct Coulomb ionization, rather than charge transfer. Moreover, because the charge of ions exceeds the target one, the perturbation theories which require $\kappa \ll 1$ and $Z_\mathrm{P} \ll Z_\mathrm{T}$ to calculate the *K*-shell ionization probability become not applicable [57].

Heavy-ion storage rings, characterized by high intense, relativistic and high-*Z* HCI beams, combined with gaseous internal targets, satisfy the experimental luminosity requirements and the isolation condition during both the collision and the decay processes [58,59] and provide new possibilities for systematic investigations of free *K*-shell hollow atoms with a wide variety of ion species, energies and charge-states. In this paper, we report the results of an x-ray spectroscopy study of *K*-shell hollow krypton atoms produced in single-collisions with 52 – 197 MeV/u bare xenon ions. The perturbation strengths cover a range from 0.70 to 1.23. The main goal of the present work is to explore the creation of *K*- and *L*-shell vacancies in krypton atoms colliding with energetic HCIs, as well as the filling process of the *K*-shell vacancies. The experiment and the analysis of the x-ray spectra are described in the next and the third sections, respectively. The mean transition energies of *K* x-rays corresponding to the outer-shell spectator vacancies are calculated and illustrated in section IV. The physical results and discussions are presented in section V, and finally a summary of the present work and a brief outlook are given in section VI.

## II. EXPERIMENT

The experiment was carried out at the HIRFL-CSR (Heavy Ion Research Facility at Lanzhou – Cooling Storage Ring) [60]. The $Xe^{27+}$ ions were produced in a superconducting electron cyclotron resonance ion source, accelerated by a sector-focusing cyclotron to 2.9 MeV/u, and then injected into the main ring (CSRm), which worked as a synchrotron in the present experiment. The ions were accumulated, cooled, and further accelerated by the CSRm to 200, 150, 100, or 60 MeV/u. On the way to the experimental ring (CSRe), the ions were stripped by a 45 mg/cm$^2$ carbon foil. Then, the bare $Xe^{54+}$ ions were selected and injected into the CSRe. The beam, which suffered energy loss in the carbon foil, was continuously cooled by an electron-cooler device [61]. The revolution frequencies were monitored using a Schottky noise analysis, and the corresponding beam energies were measured to be 197, 146, 95, and 52 MeV/u, respectively. During the experiment about $1-5\times10^7$ ions were stored. The relative momentum spread of the ions, $\Delta p/p$, was kept on a level of $2-5\times10^{-5}$. The beam energy loss due to the continuous interaction with the gas target was also compensated by the electron cooler.

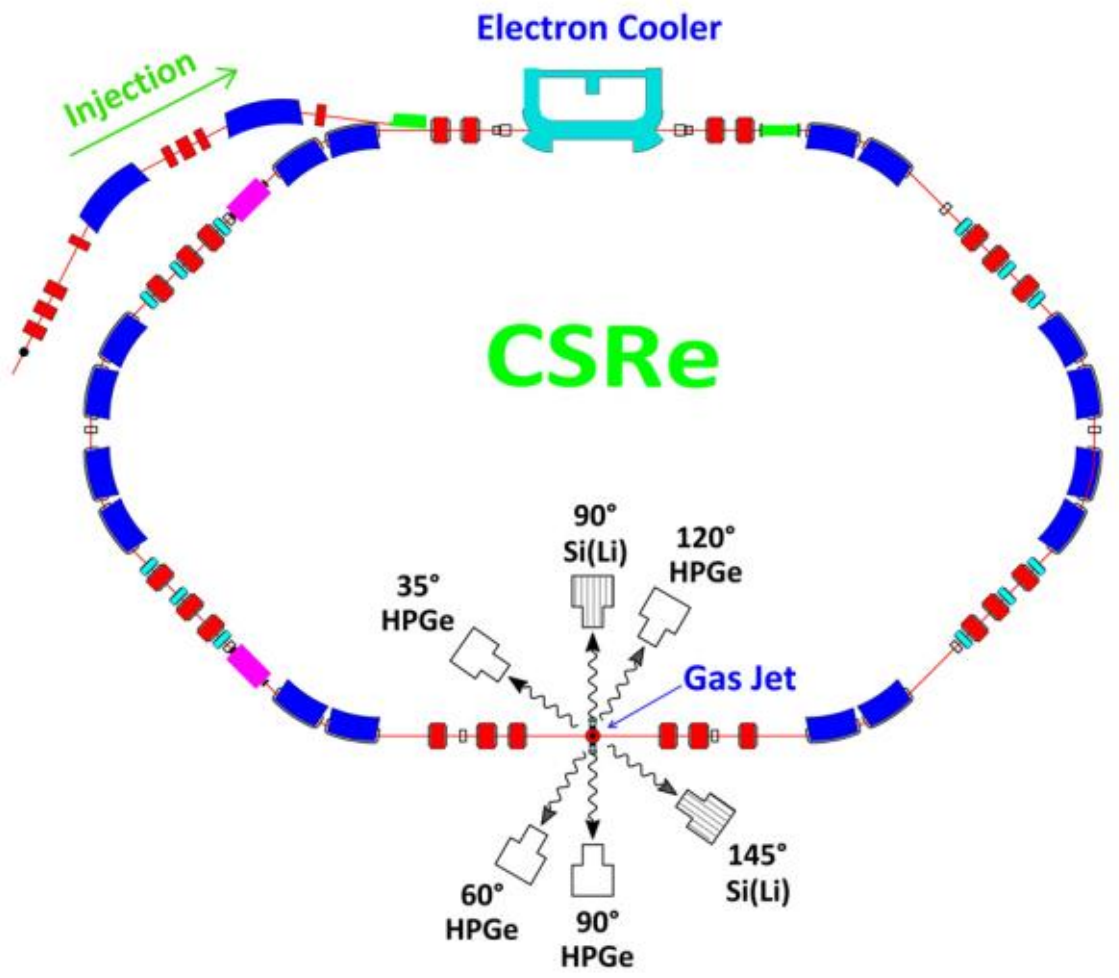

FIG. 1. (Color online) Schematic diagram of the experimental arrangement at the CSRe storage ring. Target x-rays are measured by four HPGe and two Si(Li) detectors viewing the gas-jet interaction region at angles of 35°, 60°, 90°, 120° and 145°.

The target, an atomic beam of krypton, was generated by an internal gas-jet target system [62]. It was about 3.6 mm in diameter, and had a typical thickness of $5\times10^{12}$ atoms/cm$^2$ in a background vacuum better than $2\times10^{-10}$ mbar. The overlap of the beam and target was monitored by a photomultiplier and was optimized by shifting the orbit of the ions.

The x-rays produced in collisions of the ions with the target were detected by four high-purity germanium (HPGe) detectors (ORTEC model GLP-10180/07P4) and two Si(Li) detectors (ORTEC model SLP-10180P) mounted at different observation angles with respect to the ion beam direction. A schematic of the experimental setup is shown in Fig. 1. The HPGe detectors were placed at 35°, 60°, 90° and 120°, at distances to the collision point of 500, 300, 270 and 360 mm respectively. The Si(Li) detectors were placed at 90° and 145°, at distances to the collision point of 270 and 500 mm. The detectors were separated from the ultrahigh vacuum system of the interaction chamber by 100 μm beryllium windows, shielded by lead and brass assemblies, and collimated by holes of 8 mm diameter.

The signals from the detectors were processed by standard NIM electronics, and recorded by a commercial multi-parameter multi-channel analyzer (FAST model MPA-3). The stability of the system was monitored by employing a reference pulse signal of a constant amplitude that was fed in one direction to the electronics during the experiment. The detectors were calibrated using $^{55}$Fe, $^{133}$Ba, $^{152}$Eu and $^{241}$Am radioactive sources before and after the experiment. The typical energy resolutions (FWHM) were 180 eV at 5.95 keV and 500 eV at 121 keV, respectively. The intrinsic efficiencies for the detectors were analyzed by a model introduced by Hansen *et al.* [63], where the dead layers of the detectors, the 100 μm beryllium windows of the target chamber, the 130 μm beryllium windows of the detectors, and the air between the windows were included. The detection efficiencies of the x-ray detectors were carefully calibrated, and the uncertainty of the relative efficiency in the energy region from 12 to 17 keV is estimated to be at most 3%.

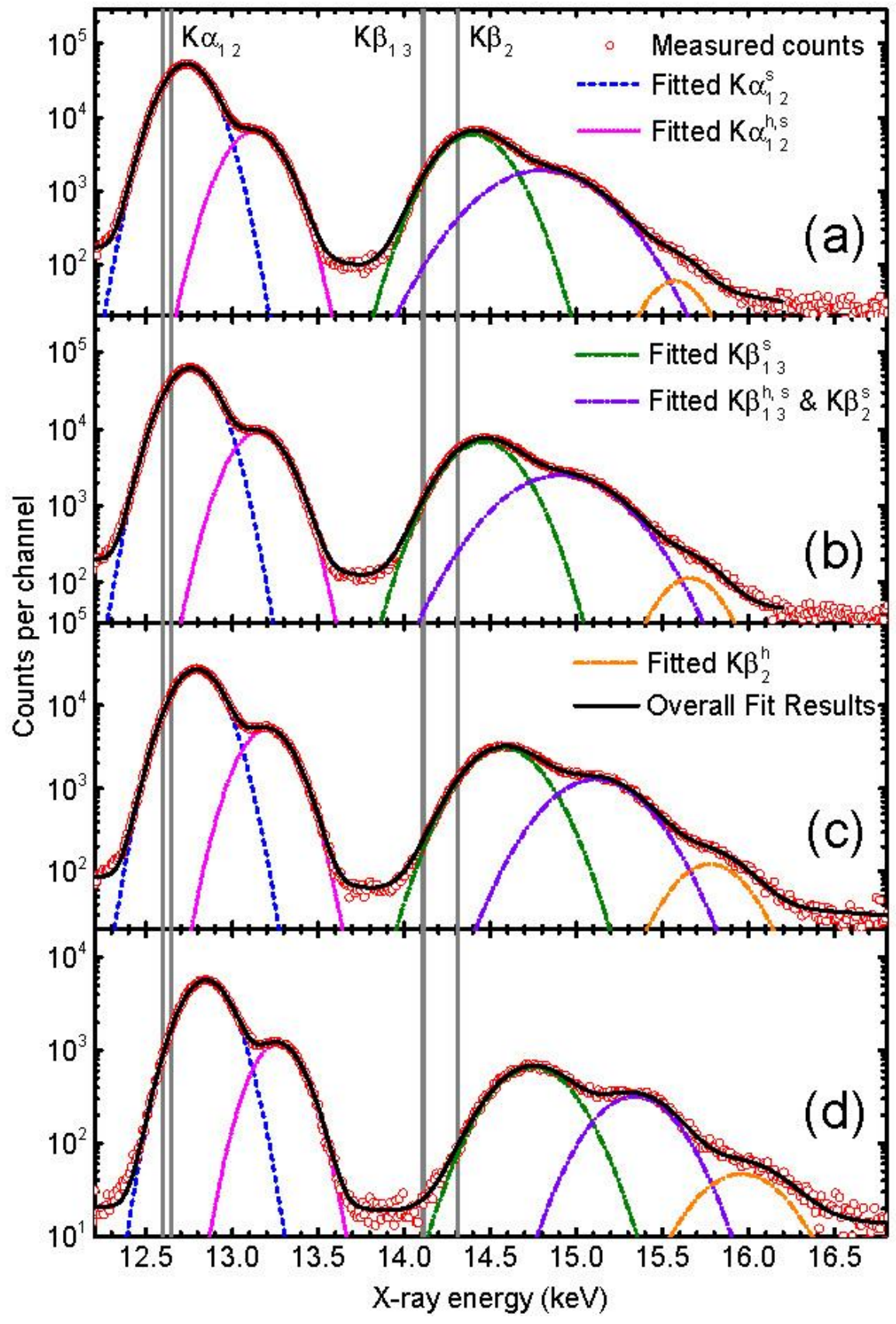

FIG. 2. (Color online) Measured spectra of x-rays emitted from krypton gas in collisions with (a) 197 MeV/u, (b) 146 MeV/u, (c) 95 MeV/u and (d) 52 MeV/u $Xe^{54+}$ ions, obtained by the Si(Li) detector at the 90° observation angle. Positions of the krypton $K\alpha_{1,2}$ diagram transitions at 12.648 and 12.595 keV, as well as the $K\beta_{1,3}$ and $K\beta_2$ transitions at 14.110 (average value) and 14.314 keV are indicated by vertical lines [64]. The measured data are represented by open circles, while the fitted peaks of $K\alpha^s_{1,2}$, $K\alpha^{h,s}_{1,2}$, $K\beta^s_{1,3}$, $K\beta^s_2$ plus $K\beta^{h,s}_{1,3}$ and $K\beta^{h,s}_2$ transitions are represented by dashed curves, respectively. The fitted background is not shown. The x-ray emitted by the projectile does not appear in this energy region at this observation angle.

The lifetime of the stored beam depended on the beam energy, and typically ranged from several minutes to tens of minutes when the target was switched on. The ions that changed their charge states during the experiment were separated from the beam using the change in their magnetic rigidity, thereby ensuring a single charge state for all of the projectile ions.

## III. X-RAY SPECTRA ANALYSIS

Typical spectra of krypton $K$ x-rays recorded by the Si(Li) detector at the 90° observation angle are shown in Fig. 2. In addition, the positions of the krypton $K\alpha_{1,2}$, $K\beta_{1,3}$ and $K\beta_2$ diagram transitions are indicated by vertical lines. Here, we denote an atomic state with $k$ $K$-shell, $l$ $L$-shell and $m$ $M$-shell vacancies as $K^{-k}L^{-l}M^{-m}$, and denote x-rays following the $K^{-1}L^{-l}M^{-m} \rightarrow K^0L^{-(l+1)}M^{-m}$ ($1 \leq l \leq 7$) transitions as $K\alpha^s$ lines, the $K^{-1}L^{-l}M^{-m} \rightarrow K^0L^{-l}M^{-(m+1)}$ ($1 \leq l \leq 8$) transitions as $K\beta^s$ lines, the $K^{-2}L^{-l}M^{-m} \rightarrow K^{-1}L^{-(l+1)}M^{-m}$ transitions as $K\alpha^{h,s}$ ($1 \leq l \leq 7$) lines, and the $K^{-2}L^{-l}M^{-m} \rightarrow K^{-1}L^{-l}M^{-(m+1)}$ ($1 \leq l \leq 8$) transitions as $K\beta^{h,s}$ lines.

The energy resolution of the present detectors in the 12 – 17 keV energy region is about 250 eV and being comparable with the roughly 50 eV energy shifts caused by each $L$-shell vacancy, leads to the $K\alpha^s_{1,2}$, $K\alpha^{h,s}_{1,2}$, $K\beta^s_{1,3}$, $K\beta^{h,s}_{1,3}$, $K\beta^s_2$ and $K\beta^{h,s}_2$ lines with different $L$-shell vacancies appear as shifted and broadened Gaussian profiles. The $K\alpha^s_{1,2}$ and $K\alpha^{h,s}_{1,2}$ peaks in the spectra are fitted by two Gaussian distributions together with a linear background. Independently, the $K\beta^s_{1,3}$, $K\beta^{h,s}_{1,3}$ plus $K\beta^s_2$ and $K\beta^{h,s}_2$ peaks are fitted by three Gaussian functions combined with another linear background. We note that the $K\beta^{h,s}_{1,3}$ and the $K\beta^s_2$ lines cannot be distinguished due to serious overlap and, therefore, are represented by a single Gaussian peak.

TABLE I. The determined energy shifts of $K\alpha^s_{1,2}$, $K\alpha^{h,s}_{1,2}$ and $K\beta^s_{1,3}$ as compared with the transitions $K^{-1}L^0 \rightarrow K^0L^{-1}$, $K^{-2}L^0 \rightarrow K^{-1}L^{-1}$ and $K^{-1}L^0M^0 \rightarrow K^0L^0M^{-1}$, as well as the relative intensities $R(K\alpha^{h,s}_{1,2}/K\alpha^s_{1,2})$ and $R(K\beta^s_{1,3}/K\alpha^s_{1,2})$ of krypton in collisions with 52 – 197 MeV/u $Xe^{54+}$ ions. The data include measurements at both the Si(Li) and the germanium detectors at the 90° observation angle. The total uncertainties in the energy shifts and the relative intensities are estimated to be within ±15 eV and ±0.002, respectively.

| Ion energy (MeV/u) | Energy shift (eV) | | | Relative intensity of X-ray emission | |
|---|---|---|---|---|---|
| | $K\alpha^s_{1,2}$ | $K\alpha^{h,s}_{1,2}$ | $K\beta^s_{1,3}$ | $\frac{K\alpha^{h,s}_{1,2}}{K\alpha^s_{1,2}}$ | $\frac{K\beta^s_{1,3}}{K\alpha^s_{1,2}}$ |
| 197 | 106 | 126 | 286 | 0.130 | 0.147 |
| 146 | 125 | 151 | 347 | 0.153 | 0.152 |
| 95 | 164 | 201 | 465 | 0.192 | 0.165 |
| 52 | 216 | 265 | 638 | 0.197 | 0.174 |

The main observed peaks of the $K\alpha$ and $K\beta$ x-rays appear to be shifted towards higher energies as compared with the corresponding diagrams lines. The hypersatellite transitions $K\alpha^{h,s}$ and $K\beta^{h,s}$ appear as shoulders on the high energy side of main $K\alpha$ and $K\beta$ peaks, and are visibly distinguished from the satellite transitions $K\alpha^s$ and $K\beta^s$.

In the fitting procedures, all of the parameters are free, in particular the widths of the peaks. The measured peak widths (FWHM) being from 300 to 600 eV, are dominated by the excitation line structures, in spite of the energy resolution of the detectors. In particular, we note that the width of the peak of $K\beta^{h,s}_{1,3}$ plus $K\beta^s_2$ varies as a result of the variation of the distance between the two groups. We also note that during the experiment, the raw data was acquired in several time segments for each energy point to test its reproducibility. The measurement accuracy of the absolute x-ray energy is within ±10 eV and the results for each segment match one other within the experimental uncertainties of several eV.

As described above, the energies and areas of the peaks have been obtained. The center of gravity energies of the $K^{-1}L^0 \rightarrow K^0L^{-1}$, $K^{-2}L^0 \rightarrow K^{-1}L^{-1}$ and $K^{-1}L^0M^0 \rightarrow K^0L^0M^{-1}$ krypton transitions are calculated from the energies and the relative probabilities of the corresponding transitions to be 12.630, 13.001 and 14.107 keV, respectively. Here both the transitions energies and probabilities are given by the GRASP 2K program, and the uncertainties of calculated energies are within ±3 eV [65,66]. The resulting energy shifts of $K\alpha^s_{1,2}$, $K\alpha^{h,s}_{1,2}$ and $K\beta^s_{1,3}$ are deduced and listed in TABLE I. The total uncertainties in the energy shift are estimated to be within ±15 eV. After calibration of the detection efficiency, the relative intensities between the $K\alpha^{h,s}_{1,2}$ and $K\alpha^s_{1,2}$ lines and the $K\beta^s_{1,3}$ and $K\alpha^s_{1,2}$ lines are presented.

It can be seen that the $K\beta$ peaks shift much more than the $K\alpha$ peaks. This is due to the fact that the effect of additional $L$-vacancies on $M$-shell orbits is greater than that on $L$-shell orbits. In addition, the hypersatellite lines shift more than the corresponding satellite lines when the projectile energy is same. Furthermore, the lower the projectile energy, the greater the X-ray energy shift and the higher the ratio $R(K\alpha^{h,s}_{1,2}/K\alpha^s_{1,2})$, although the later varies much more slowly than the former.

## IV. ENERGY CALCULATION OF $K\alpha^s_{1,2}$, $K\alpha^{h,s}_{1,2}$ AND $K\beta^s_{1,3}$ LINES

When an atom is impacted by an energetic heavy ion, the production of $K$-shell vacancies is usually accompanied by the creation of other vacancies in higher shells. The production of such $K$-shell hollow atoms containing higher-shell vacancies is a unique feature of heavy-ion collisions. In the present work, the relative probabilities for production of the $K$-hollow krypton atoms with respect to single $K$-shell ionized ones can be deduced directly from relative intensities of $K\alpha^s_{1,2}$ and $K\alpha^{h,s}_{1,2}$ after taking into account their respective fluorescence yields. The configurations of $L$- and $M$-shell vacancies, which also is required for evaluation of the fluorescence yields of $K$ x-rays, have to be inferred from the energy shifts of the $K\alpha^{s}_{1,2}$, $K\beta^{s}_{1,3}$ and $K\alpha^{h,s}_{1,2}$ with respect to corresponding x-rays from full $L$- and $M$-shells [64,67].

Hence, we have calculated the mean transition energies of $K\alpha^{s}_{1,2}$, $K\alpha^{h,s}_{1,2}$ and $K\beta^{s}_{1,3}$ corresponding to the outer-shell spectator vacancies using the GRASP 2K program [65,66], as shown in Fig. 3. Two extreme cases: a full and an empty $M$-shell in the final state are illustrated. These results clearly show that the energy shift is proportional to the $L$-vacancy number in each case. Therefore, we assume that the energy shift caused by the $M$-vacancy is also proportional to its number. In this approximation, an $M$-shell vacancy shifts the $K\alpha^{s}_{1,2}$ and $K\alpha^{h,s}_{1,2}$ transitions to higher energies about 5 – 8 eV, and shifts the $K\beta^{s}_{1,3}$ transitions about 18 – 30 eV, respectively. More significantly, an $L$-shell vacancy shifts them by about 45 – 56 and 100 – 130 eV, respectively. Taking into account the electron numbers in each shell, the whole $M$-shell shifts the $K$ x-ray transition energy by roughly 1/3 as much as the whole $L$-shell, and so is not negligible. Nevertheless, the energy shifts of $K\alpha^{s}_{1,2}$, $K\alpha^{h,s}_{1,2}$ and $K\beta^{s}_{1,3}$ transitions of a state with $l$ $L$- and $m$ $M$-shell vacancies, $\Delta E_{l,m}$, can be obtained.

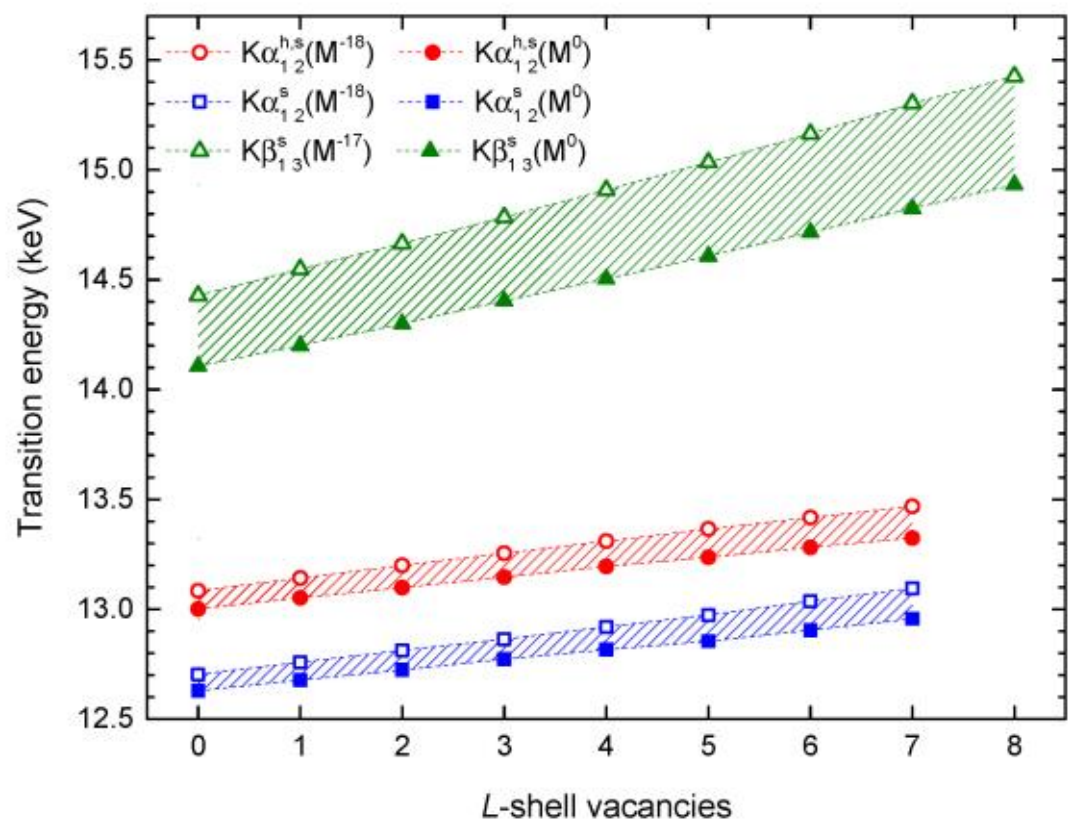

FIG. 3. (Color online) Transition energy of krypton $K\alpha^{s}_{1,2}$, $K\alpha^{h,s}_{1,2}$ and $K\beta^{s}_{1,3}$ versus the number of the spectator vacancies in the $L$-shell, calculated using the GRASP 2K program [65,66]. Two extreme cases are illustrated: a full and an empty $M$-shell (final states).

In close collisions with energetic heavy ions, i.e., when at least one $K$-shell vacancy is created, the population distribution of the number of the $L$- and $M$-shell vacancies in target atoms, $l$ and $m$, are both expected to be approximately binomial [68], and are denoted here by $p^{L}_{0}(l)$ and $p^{M}_{0}(m)$, respectively. Because vacancy rearrangement may happen prior to the $K$ x-ray emission, a set of new distributions $p^{L}_{x}(l)$ and $p^{M}_{x}(m)$ at the $K$ x-ray emission time, which could be slightly different from $p^{L}_{0}(l)$ and $p^{M}_{0}(m)$, should be introduced to describe the state distribution when a $K$ x-ray is emitted. It also should be noted that the measured $K\alpha^{s}_{1,2}$ and $K\beta^{s}_{1,3}$ lines may originate either from single $K$-shell ionization events or from cascade processes of double $K$-shell ionization events. As will be shown latter, the average outer-shell spectator vacancies of the latter events are larger than the former ones. But we do not try to separate these two groups of distributions of $l$ and $m$ considering the present experimental resolution. Moreover, $p^{L}_{x}(l)$ and $p^{M}_{x}(m)$ are also expected to be approximately binomial to simplify the equations. The distributions are characterized by the average vacancy numbers in the $L$- and $M$-shells when a $K$ x-ray is emitted, $\bar{l}_x$ and $\bar{m}_x$, respectively.

$$p^{L}_{x}(\bar{l}_x; l) = \binom{8}{l} \left(\bar{l}_x/8\right)^{l} \left(1-\bar{l}_x/8\right)^{8-l}, \tag{1}$$

and

$$p^{M}_{x}(\bar{m}_x; m) = \binom{18}{m} (\bar{m}_x/18)^{m} (1-\bar{m}_x/18)^{18-m}, \tag{2}$$

where $0 \le l \le 8$, $0 \le m \le 18$. Therefore, by taking into account the fluorescence, the mean energy shift of those states with average $\bar{l}_x$ and $\bar{m}_x$ vacancy in $L$- and $M$-shell are

$$\overline{\Delta E}_{K\alpha} = \frac{\sum_{l=0}^{7}\sum_{m=0}^{18} p^{L}_{x}(\bar{l}_x;l)\, p^{M}_{x}(\bar{m}_x;m)\, \omega_{K\alpha}(l,m)\, \Delta E_{l,m}}{\sum_{l=0}^{7}\sum_{m=0}^{18} p^{L}_{x}(\bar{l}_x;l)\, p^{M}_{x}(\bar{m}_x;m)\, \omega_{K\alpha}(l,m)}, \tag{3}$$

for $K\alpha^{s}_{1,2}$ and $K\alpha^{h,s}_{1,2}$ transitions, and

$$\overline{\Delta E}_{K\beta} = \frac{\sum_{l=0}^{8}\sum_{m=0}^{17} p^{L}_{x}(\bar{l}_x;l)\, p^{M}_{x}(\bar{m}_x;m)\, \omega_{K\beta}(l,m)\, \Delta E_{l,m}}{\sum_{l=0}^{8}\sum_{m=0}^{17} p^{L}_{x}(\bar{l}_x;l)\, p^{M}_{x}(\bar{m}_x;m)\, \omega_{K\beta}(l,m)}, \tag{4}$$

for $K\beta^{s}_{1,3}$, respectively.

The average vacancy numbers in $L$- and $M$-shells produced in the collisions, $\bar{l}_0$ and $\bar{m}_0$ can be related to each other via a universal scaling formula of the geometrical model developed by Sulik *et al* [69]. In this model, the simultaneous inner and outer shell ionization processes are characterized by the inner shell ionization cross section and the ionization probability per electron for the outer shells at zero impact parameter within the framework of the independent electron approximation (IPM). Hence, the mean ionization probability per electron for both $L$- and $M$-shell shells could be described by a universal scaling parameter $X$. Specifically,

$$\bar{l}_0 = \frac{8X_L^2}{4.2624+X_L^2\left(1+0.5e^{-X_L^2/16}\right)}, \tag{5}$$

and

$$\bar{m}_0 = \frac{18X_M^2}{4.2624+X_M^2\left(1+0.5e^{-X_M^2/16}\right)}, \tag{6}$$

where $X_{L,M} = 4\alpha c Z_{\mathrm{P}} V_{L,M}\sqrt{G\left(V_{L,M}\right)/n v_{\mathrm{P}}}$ is a universal variable. In this expression, $\alpha$ is the fine structure constant, $c$ is the speed of light, $Z_{\mathrm{P}}$ is the nuclear charge of the ion and $v_{\mathrm{P}}$ is its speed, $V_{L,M} = v_{\mathrm{P}}/v_{L,M}$ while $v_{L,M}$ is the classical speed of the target electrons, $n = 2$ for the $L$-shell and $n = 3$ for the $M$-shell, respectively. In the present work, $v_{L,M}$ is derived from the $L$- or $M$-shell average binding energy of krypton, and the value of $G\left(V_{L,M}\right)$ is calculated using the binary encounter approximation (BEA) scaling function of Gryzinski [70]. As a result, an approximation relation of

$$\bar{m}_0 \approx 1.4\bar{l}_0 + 0.093\bar{l}_0^2, \tag{7}$$

is obtained. If vacancy rearrangement before a $K$ x-ray decay does not significantly change its distribution, which can be expected in the case of heavy atoms, $\bar{m}_x$ and $\bar{l}_x$ then satisfy the same relation. It should be noted that the relation between the average number of $L$- and $M$-shells can also be estimated from the measured intensity ratio $I_{K\beta^s_{1,3}}/I_{K\alpha^s_{1,2}}$ [71]. If the number of $L$-vacancies obtained via different estimation methods is set to be equal, then the number difference in the vacancy in $M$-shells is at most one. Since the shifts due to $M$-vacancies are generally about an order of magnitude smaller than those due to $L$-shells, the deduced number of vacancies in $L$-shells is not strongly affected.

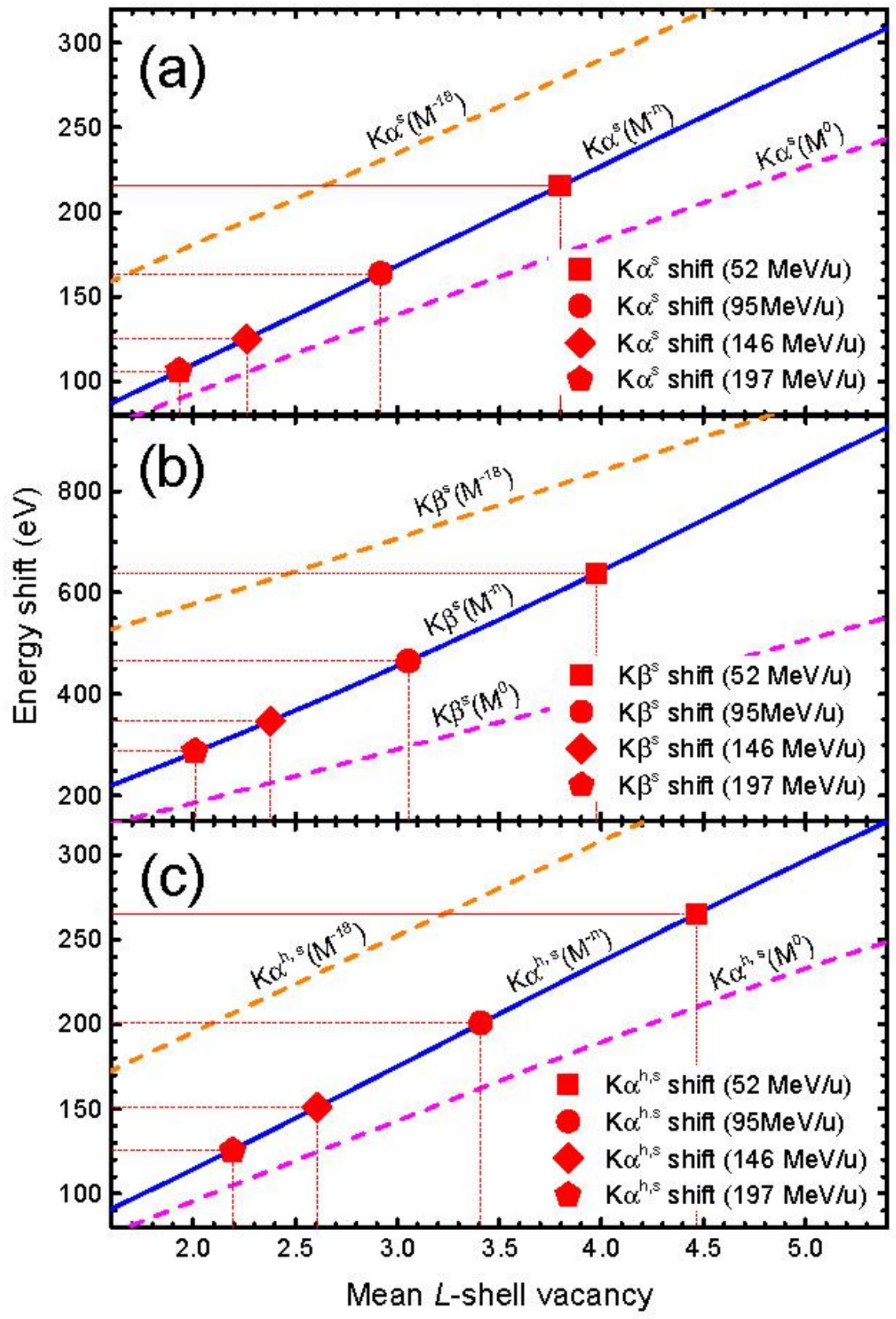

FIG. 4. (Color online) Energy shift of (a) $K\alpha^s_{1,2}$, (b) $K\beta^s_{1,3}$ and (c) $K\alpha^{h,s}_{1,2}$ lines versus mean $L$-shell vacancy number in krypton. The solid blue lines represent a partially ionized $M$-shell according to formula (7), while an empty or a full $M$-shell is represented by the dashed lines. The experimental datasets in collisions with $Xe^{54+}$ ions of 52, 95, 146 and 197 MeV/u are represented by solid squares, circles, diamonds and pentagons, respectively. The mean $L$-shell vacancy numbers are indicated by vertical lines.

The fluorescence for the double-$K$-vacancy state of krypton has been calculated by Chen [72]. However, most fluorescence yields of multi-vacancy krypton states are currently not available. Therefore, fluorescence yields of multi-vacancy configurations $\omega_{K\alpha^s_{1,2}}(l,m)$, $\omega_{K\alpha^{h,s}_{1,2}}(l,m)$ and $\omega_{K\beta^s_{1,3}}(l,m)$ are derived using a statistical weighting procedure developed by Larkins *et al.* [73]. In this procedure, the original various radiative and radiationless transition rates of a single $K$-vacancy are obtained from tabulated data [74,75], and then the multi-vacancy transition rates are scaled according to the vacancy configuration.

Finally, the calculated mean energy shifts of the $K\alpha^s_{1,2}$, $K\alpha^{h,s}_{1,2}$ and $K\beta^s_{1,3}$ lines versus the average numbers of $L$-shell vacancies $\bar{l}_x$ are obtained and plotted in Fig. 4. In the figure the experimental data are also shown, where the energy shifts and the mean $L$-shell vacancies are indicated by horizontal and vertical lines, respectively.

## V. RESULTS AND DISCUSSION

### A. Average number of spectator *L*-vacancies during the *K* x-ray emission

Using the calculated results of energy shifts with the average numbers of $L$-shell vacancies described in Sec. IV, the experimental value $\bar{l}_x$ for the present work were extracted from determined energy shifts and listed in TABLE II. Actually, both lines $K\alpha^s_{1,2}$ and $K\beta^s_{1,3}$ include contributions from both single and double $K$-shell ionization events. The discrepancy between the two data sets is within 10%, this result confirms the reliabilities of the present method because both the $K\alpha^s_{1,2}$ and the $K\beta^s_{1,3}$ lines reflect the same $L$-vacancy configuration of single $K$-vacancy atom. The values deduced from the $K\alpha^{h,s}_{1,2}$ lines include only contributions from double $K$-shell ionization events.

TABLE II. Deduced mean $L$-vacancies $\bar{l}_x$ and average fluorescence yield ratios $\bar{\omega}_{K\alpha^{h,s}_{1,2}}/\omega_{K\alpha_{1,2}}$, $\bar{\omega}_{K\alpha^s_{1,2}}/\omega_{K\alpha_{1,2}}$ and $\bar{\omega}'_{K\alpha^s_{1,2}}/\omega_{K\alpha_{1,2}}$ of krypton in collisions with 52 – 197 MeV/u $Xe^{54+}$ ions. The uncertainty of the deduced mean $L$-vacancies is estimated to be ±0.26.

| ion energy (MeV/u) | mean $L$-vacancies | | | fluorescence yields ratio | | |
|---|---|---|---|---|---|---|
| | from $K\alpha^s_{1,2}$ | from $K\beta^s_{1,3}$ | from $K\alpha^{h,s}_{1,2}$ | $\frac{\bar{\omega}_{K\alpha^s_{1,2}}}{\omega_{K\alpha_{1,2}}}$ | $\frac{\bar{\omega}_{K\alpha^{h,s}_{1,2}}}{\omega_{K\alpha_{1,2}}}$ | $\frac{\bar{\omega}'_{K\alpha^s_{1,2}}}{\omega_{K\alpha_{1,2}}}$ |
| 197 | 1.93 | 2.01 | 2.19 | 1.081 | 1.093 | 1.136 |
| 146 | 2.26 | 2.38 | 2.61 | 1.096 | 1.111 | 1.153 |
| 95 | 2.92 | 3.05 | 3.41 | 1.125 | 1.145 | 1.175 |
| 52 | 3.80 | 3.97 | 4.46 | 1.160 | 1.176 | 1.152 |

Two remarkable features should be noted. First, our results show that with increasing the projectile energy the mean $L$- and $M$-shell vacancies values accompanying with the $K$-shell ionization decrease. The collision parameters is sufficiently small and the collision energy is sufficiently high for the $L$- and $M$-shell electrons, accompanying with $K$-shell ionization in the present case. Therefore, this phenomenon could be qualitatively understood with the classical pictures of ionization and capture processes [76], according to which, both the ionization and capture probabilities decrease with increasing of collision energies. Second, when a hypersatellite transition occurs, there are systematically more spectator vacancies than in the case of a satellite transition. This implies that an increasing quantity of the $K$-shell ionization by decreasing the impact parameter coexists with an increasing the mean $L$-shell vacancy creation. This result is consistent with that of Horvat *et al.*, where the $K\alpha^s_{1,2}$ and $K\alpha^{h,s}_{1,2}$ lines emitted from argon gas at atmospheric pressure under bombardment by 10 MeV/u heavy ions were measured [46] . As mentioned before, a small fraction of the satellite transitions is a part of cascade decays of the double *K*-shell ionized states, but this does not interfere with our conclusion.

Accordingly, these fluorescence yields, which depend on $\bar{l}_x$ and $\bar{m}_x$, are obtained using the statistical weighting procedure [73] mentioned above and averaging over the target vacancy configuration distribution. The calculated relative fluorescence yields are listed in TABLE II. Here $\omega_{K\alpha_{1,2}}$ is the fluorescence yield of the $K\alpha_{1,2}$ line originated from the single-*K*-vacancy state without spectator *L*-vacancies, and the $\bar{\omega}_{K\alpha_{1,2}^{h,s}}$, $\bar{\omega}_{K\alpha_{1,2}^{s}}$ and $\bar{\omega}'_{K\alpha_{1,2}^{s}}$ in the table are average fluorescence yields of the $K\alpha_{1,2}^{h,s}$ line originated from the $K^{-2}$, the $K\alpha_{1,2}^{s}$ line originated from the $K^{-1}$, and the $K\alpha_{1,2}^{s}$ line originated from the cascade decay of the $K^{-2}$ states, respectively. Since the $\bar{l}'_x$ of the $K^{-1}$ states originated from the $K^{-2}$ states are not deducible from experimental results, the values are taken to be $\bar{l}_x(K\alpha_{1,2}^{h,s})+1$ in accordance with the rough approximation that the dominant *K*-*L* radiative and *K*-*LM* auger decay [74,75] while filling the first *K*-shell vacancy will both increase the vacancy number in the *L*-shell by one.

The result shows that the relative fluorescence yields of $K\alpha$ for krypton varies slowly with respect to $\bar{l}_x$ when it is less than 5.5. In addition, the vacancy rearrangements in the *L*-shell almost do not change this ratio of average fluorescence yields (i.e., $\bar{\omega}_{K\alpha_{1,2}^{h,s}}/\bar{\omega}_{K\alpha_{1,2}^{s}}$ and $\bar{\omega}_{K\alpha_{1,2}^{h,s}}/\bar{\omega}'_{K\alpha_{1,2}^{s}}$) because the filling of the *K*-shell vacancy is much faster than that of *L*-shell vacancy as the *L*-shell Coster-Kronig transitions are usually energetically forbidden in multiply ionized krypton atoms [77,78].

### B. Ratio of *K*-shell hollow to singly ionized atoms

The cross-section ratio between double and single *K*-shell ionization, $R_{21}$, can be calculated from

$$R_{21} \equiv \frac{\sigma_{K^{-2}}}{\sigma_{K^{-1}}} = \frac{N_{K\alpha_{1,2}^{h,s}}/\bar{\omega}_{K\alpha_{1,2}^{h,s}}}{N_{K\alpha_{1,2}^{s}}/\bar{\omega}_{K\alpha_{1,2}^{s}} - N_{K\alpha_{1,2}^{h,s}}/\bar{\omega}'_{K\alpha_{1,2}^{s}}}$$

$$= \frac{R_{21}^{x}}{\bar{\omega}_{K\alpha_{1,2}^{h,s}}/\bar{\omega}_{K\alpha_{1,2}^{s}} - R_{21}^{x}\bar{\omega}_{K\alpha_{1,2}^{h,s}}/\bar{\omega}'_{K\alpha_{1,2}^{s}}}, \quad (8)$$

where $N_{K\alpha_{1,2}^{h,s}}$, $N_{K\alpha_{1,2}^{s}}$ are photon counts in the experiment, then $R_{21}^{x}$ are the ratio between them. By combining the relative intensities of x-ray emission listed in TABLE I and relative fluorescence yields listed in TABLE II, the cross-section ratio $R_{21}$ for various collision energy were obtained and plotted versus the factor $\kappa$ in Fig. 5. Phenomenally, it shows that the cross-section ratio $R_{21}$ increases linearly with $\kappa$ in the smaller region and less rapidly at the larger region, and the turning point seems near $\kappa \sim 1$.

The target *K*-shell ionization by light energetic ions is usually described by various perturbative approaches, e.g., the plane wave Born approximation (PWBA) [57], the semi-classical approximation (SCA) [79,80] and the binary encounter approximation (BEA) [81,82]. When the perturbation strength $\kappa$ is small, the single *K*-shell ionization cross section is roughly proportional to $\kappa^2$ [35,36], while the ionization of the two *K*-shell electrons can be attributed to two successive independent collisions with the projectile in the frame of independent electron model, i.e., the so-called "two-step" mechanism. Consequently, the double-ionization cross section is proportional to $\kappa^4$ [21,35,83]. In the present work, the velocity of the $Xe^{54+}$ ions with energy of 197, 146, 95 and 52 MeV/u is 77, 69, 57, and 44 *a.u.*, and the corresponding perturbation strength $\kappa$ is 0.70, 0.78, 0.94 and 1.23, respectively. The experimental results show that the cross-section ratio increases linearly with $\kappa$, rather than $\kappa^2$, and therefore confirms that the perturbative approaches break down when $\kappa$ is comparable to unity [57]. The non-perturbative methods, such as the Magnus approximation [84] and the Glauber approximation [85-87], predict that the ionization cross section grows slowly than $\kappa^2$ when $\kappa \sim 1$, which qualitatively agrees with the present results.

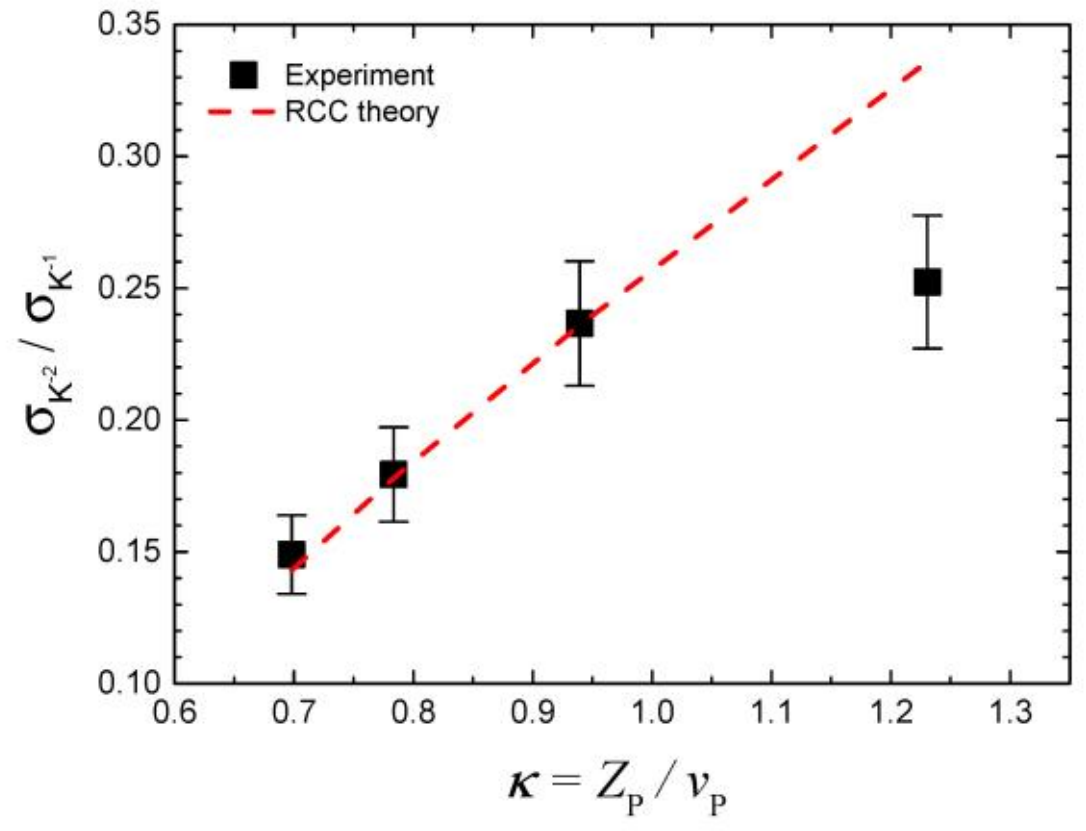

FIG. 5. (Color online) Cross-section ratio between double and single *K*-shell ionization of krypton in collisions with 52 – 197 MeV/u $Xe^{54+}$ ions. The horizontal coordinate is the perturbation strength $\kappa = Z_P/v_P$, where $Z_P = 54$ is the charge of the projectile and $v_P$ is its velocity in atomic units. The experimental data are represented by the solid squares. The theoretical results of the relativistic coupled-channel method are represented by the dashed line.

Recently, Kozhedub *et al.* developed a relativistic coupled-channel (RCC) method based on independent electron model and two-center atomic-like Dirac-Fock-Sturm orbitals as a basis set [88]. The method allows one to unperturbatively describe the relativistic quantum dynamics of electrons in ion-atom collisions, including the *K*-shell-to-*K*-shell electron transfer channel. Preliminary calculations by this method are also shown in Fig. 5. The theory agrees with the present experiment very well at the higher energies, especially the linear dependence of $R_{21}$ on $\kappa$ as well as the right slope when $\kappa < 1$. But it fails to reproduce the experiment data when $\kappa > 1$, and the reason of this deviation is not clear yet.

## VI. SUMMARY AND OUTLOOK

The $K\alpha_{1,2}^{s}$, $K\alpha_{1,2}^{h,s}$, $K\beta_{1,3}^{s}$ and $K\beta_{1,3}^{h,s}$ x-rays of krypton in collisions with 52 – 197 MeV/u $Xe^{54+}$ ions have been measured. The relative yield of *K*-hollow krypton atoms with respect to single *K*-shell ionized ones was determined to be as high as 14 – 24%. In our previous work of 185 MeV/u $Ni^{19+}$-Kr collisions [89] no krypton hypersatellite lines were observed. Our work confirms that the charge state of the projectile ions plays a dominant role in production of *K*-shell hollow krypton in the present energy region. Different from the previous work of $Fe^{26+}$ with argon collisions at 7.7 MeV/u [39], in which a high relative yield of *K*-hollow atoms was created mainly by the *K*-shell-to-*K*-shell electron transfer, the present dominant *K*-vacancy creation mechanism is due to the direct Coulomb ionization in the non-perturbative regime.

The mean spectator *L*-vacancies were estimated from the energy shifts of the transitions. In particular, it is shown that for the higher the projectile energy less additional vacancies are created in the target *L*-shells accompanying the *K*-shell ionizations. We also observed that more spectator *L*-vacancies are created accompanying double *K*-shell ionization than single one. The measured double-to-single *K*-shell ionization cross-section ratio is proportional to the perturbation strength $\kappa$ in the rough region of $\kappa < 1$, but shows less rapidly increasing when $\kappa > 1$. It implies the breakdown of the first order perturbative approaches. The present experimental result is compared with a preliminary calculation of the RCC method. The theory reproduced the present experiment very well in the region of $\kappa < 1$, but unexpectedly deviated from the experimental data when $\kappa > 1$. In order to clarify this deviation, further experimental and non-perturbative theoretical studies are urgently demanded.

The present work shows that a heavy ion storage ring, equipped with an internal gas-jet target, is an efficient setup to produce free and heavy *K*-shell hollow atoms. By utilizing different kinds of projectile ions, *K*-shell hollow atoms with different additional *L*-vacancies could be investigated systematically. With the new generation of storage rings which will be available in the near future [90,91], the ion beam will be three orders of magnitude stronger than present ones, and thus factories to produce a large amount of free and heavy *K*-shell hollow atoms can be expected. Combined with high-resolution, but much smaller observed solid angle, X-ray spectroscopy techniques, e.g., crystal spectrometers and micro-calorimeters, exotic atoms with several hollow or open inner shells, as well as exotic transitions involving more than one electron and one photon in a strong field (e.g., two-electron-one-photon transitions, double-photon transitions, etc.) may be systematically explored.

## ACKNOWLEDGMENT

We thank the crew of the accelerator department for their hardworking operation of the HIRFL-CSR complex. This work was supported by the National Natural Science Foundation of China under Grants Nos. U1332206, 11179017, 11275240, 11375201 and U1532130. YSK acknowledges the support from RFBR-NSFC (Grant No. 17-52-53136).

+=+=+============+=+=+